\documentclass[fleqn,usenatbib]{mnras}

\usepackage{newtxtext,newtxmath}
\usepackage[T1]{fontenc}
\DeclareRobustCommand{\VAN}[3]{#2}
\let\VANthebibliography\thebibliography
\def\thebibliography{\DeclareRobustCommand{\VAN}[3]{##3}\VANthebibliography}

\usepackage{graphicx}	
\usepackage{amsmath}	

\newcommand\bb[1]{\mbox{\boldmath{$#1$}}}
\newcommand\grad{\bb{\nabla}}
\newcommand\bcdot{\,\bb{\cdot}\,}

\newcommand\btimes{\,\bb{\times}\,}
\newcommand{\vecv}{\bb{v}}
\newcommand{\vecB}{\bb{B}}
\newcommand{\vecu}{\bb{u}}

\newcommand{\upar}{u_{\|}}
\newcommand{\uperp}{u_\perp}

\newcommand{\vecR}{\bb{R}}
\newcommand{\vecE}{\bb{E}}
\newcommand{\Epar}{E_{\|}}
\newcommand{\vecvcurv}{\bb{v}_\mathrm{curv}}
\newcommand{\vecvpol}{\bb{v}_\mathrm{pol}}
\newcommand{\vecvgradB}{\bb{v}_{\boldsymbol{\nabla}B}}
\newcommand{\vecvrel}{\bb{v}_\mathrm{rel}}
\newcommand{\apar}{a_{\|}}
\newcommand{\acurv}{a_{\mathrm{curv}}}
\newcommand{\agradB}{a_{\boldsymbol{\nabla}B}}

\newcommand{\vecb}{\bb{b}}
\newcommand{\rmd}{\mathrm{d}}
\DeclareMathAlphabet\mathbfcal{OMS}{cmsy}{b}{n}

\title[Particles in Turbulent Coronal Loops]{Particle Trapping and Acceleration in Turbulent Post-flare Coronal Loops}
\author[F.\ Bacchini et al.]{
Fabio Bacchini$^{1,2}$\thanks{E-mail: fabio.bacchini@kuleuven.be},
Wenzhi Ruan$^{1}$, and Rony Keppens$^{1}$
\\
$^{1}$Centre for mathematical Plasma Astrophysics, Department of Mathematics, KU Leuven, Celestijnenlaan 200B, B-3001 Leuven, Belgium\\
$^{2}$Royal Belgian Institute for Space Aeronomy, Solar-Terrestrial Centre of Excellence, Ringlaan 3, 1180 Uccle, Belgium
}

\begin{document}
\label{firstpage}
\pagerange{\pageref{firstpage}--\pageref{lastpage}}
\maketitle

\begin{abstract}
We present a study of energetic-electron trapping and acceleration in the Kelvin--Helmholtz-induced magnetohydrodynamic (MHD) turbulence of post-flare loops in the solar corona. Using the particle-tracing capabilities of \texttt{MPI-AMRVAC 3.0}, we evolve ensembles of test electrons (i.e.\ without feedback to the underlying MHD) inside the turbulent looptop, using the guiding-center approximation. With the MHD looptop model of \cite{Ruan2018}, we investigate the relation between turbulence and particle trapping inside the looptop structure, showing that better-developed turbulent cascades result in more efficient trapping primarily due to mirror effects. We then quantify the electron acceleration in the time-evolving MHD turbulence, and find that ideal-MHD processes inside the looptop can produce nonthermal particle spectra from an initial Maxwellian distribution. Electrons in this turbulence are preferentially accelerated by mirror effects in the direction perpendicular to the local magnetic field while remaining confined within small regions of space between magnetic islands. Assuming dominance of Bremsstrahlung radiation mechanisms, we employ the resulting information from accelerated {electrons} (combined with the MHD background) to construct HXR spectra of the post-flare loop that include nonthermal-particle contributions. Our results pave the way to constructing more realistic simulations of radiative coronal structure for comparison with current and future observations.
\end{abstract}

\begin{keywords}
acceleration of particles -- Sun: corona -- Sun: flares -- Sun: X-rays, gamma-rays -- turbulence
\end{keywords}




\section{Introduction}
\label{sec:intro}

\begin{figure}
\centering
\includegraphics[width=1\columnwidth, trim={0mm 0mm 0mm 0mm}, clip]{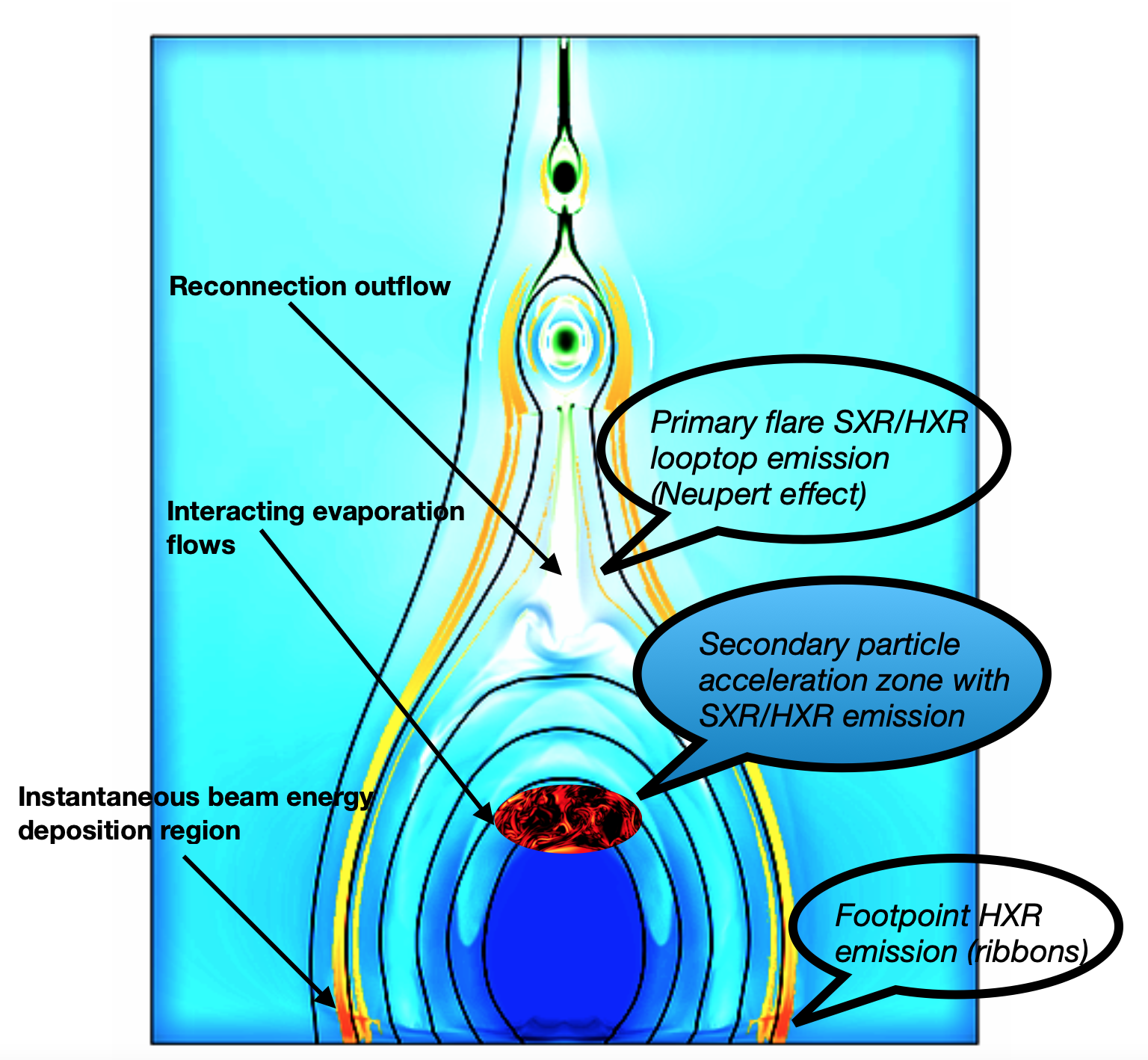}
\caption{Schematic representation of a flaring loop, using the self-consistent interaction between thermal plasma (MHD) and accelerated energetic beams (in orange), from \protect\cite{druett2023AA}. This work focuses only on the central region (see blue balloon), where secondary particle acceleration and SXR/HXR emission may occur due to rising flows meeting and mixing deep inside the looptop.}
\label{fig:looptop_sketch}
\end{figure}

Solar flares are sudden brightening phenomena that occur frequently in the solar atmosphere. Local broadband emission (e.g.\ EUV, soft X-ray, and hard X-ray) increases significantly when a solar-flare event occurs, and the required energy for the emission is likely sourced by the solar magnetic field. {A classical explanation of this mechanism (\citealt{Shibata1995,Priest2002}; and \citealt{Fang2016} for the last point below) is described in several steps: (1) magnetic reconnection in the region between a coronal loop and a suspended flux rope above causes a rapid release of magnetic energy at coronal height; (2) the configuration of the local magnetic field is altered and magnetic arcades form, the footpoints of which are at the solar surface; (3) the released energy is then transported downward toward the chromosphere; (4) energy deposited at the chromosphere causes upward evaporation flows, which fill the magnetic arcades with hot, dense plasma; (5) the evaporation flows deep inside the loop can mix into a highly turbulent state, creating chaotic distributions of magnetic fields threading hot plasma. A schematic representation of this process is shown in Fig.~\ref{fig:looptop_sketch}, which is in fact a cartoonized variant of a recent multidimensional MHD simulation which incorporates dynamically and self-consistently the role of particle beams (visible in orange) in the entire flare system (\citealt{druett2023AA}). In this work, we specifically focus on the inner region (blue balloon in Fig.~\ref{fig:looptop_sketch}).}

Hot plasma in coronal loops releases strong EUV and soft-X-ray (SXR) emission, bright enough for loops to frequently appear in EUV and SXR images of flare events (e.g.\ \citealt{Su2013,Nindos2015,Chen2017}). Separated hard-X-ray (HXR) sources can often be found in the flaring region, particularly at the footpoints of the EUV and SXR loops and near the looptop (e.g.\ \citealt{Masuda1994,chen2020NatAs}), suggesting that energetic electrons were produced during the events. The looptop region can potentially source radiation from processes that are secondary to the main impulsive reconnection described above. This secondary generation of radiation has been relatively underexplored, particularly in terms of the associated dynamics of energetic particles emitting in the X-ray wavelengths.

In general, the generation of energetic {electrons} is a hot topic in the study of solar flares, as particle acceleration and the following collisional energy loss act as an important path of energy transfer and transport (e.g.\ \citealt{Kerr2020}). Energetic electrons are thought to be generated at coronal height, produce looptop HXR source(s), to then move to the chromosphere along magnetic-field lines and produce footpoint sources, losing most of their energy there through collisions (\citealt{Krucker2008,Benz2017}). During a solar-flare event, up to $10^{32}$~erg of energy is released via magnetic reconnection, and up to 50\% of the released energy is involved in the generation of nonthermal electrons (e.g.\ \citealt{Aschwanden2017}). Potential acceleration mechanisms include electric DC-field acceleration, stochastic acceleration, and shock acceleration (e.g.\ \citealt{Zharkova2011}). The latest so-called ``self-consistent'' models for standard flare scenarios do not (yet) address the detailed acceleration aspects, but do allow for two-way coupling between the thereby generated energetic electron beams and the full thermal plasma evolutions (\citealt{ruan2020,druett2023AA,druett2023SP}). Within such models, the post-flare loops underneath the current sheet show clear evaporations, which in these models can be due to thermal conduction, reflection and/or be entirely beam-driven.

\cite{Fang2016} proposed a scenario for the generation of the nonthermal electrons in which evaporation from the chromosphere produces turbulence at coronal height via the Kelvin--Helmholtz instability (KHI), and electrons are then accelerated in the turbulence through stochastic processes. In that scenario, KHI turbulence may also confine the accelerated electrons inside the looptop, concurring in the creation of the strong looptop HXR sources that have been observed. In observations, strong chromospheric evaporation at hundreds of km/s is frequently found in flare events (e.g.\ \citealt{Milligan2009,Tian2014}). Magnetohydrodynamic (MHD) simulations demonstrate that these fast evaporation events have the ability to produce KHI turbulence inside flare loops (e.g.\ \citealt{Fang2016,Ruan2018,Ruan2019}). The contribution of KHI turbulence to the production of solar-flare nonthermal electrons (as opposed to direct acceleration in the current sheet/reconnection site above) remains to be assessed. It is relatively difficult for such a multistep process to efficiently transfer (i.e., up to a fraction of 50\%) the reconnection-released energy to nonthermal electrons (\citealt{Cargill1996,Miller1997}). Nevertheless, some fraction of the nonthermal electrons may still be produced in this way, but this has not been quantitatively verified in simulations.

To study the energization of particles in the solar corona, particle methods are often employed in simulations to obtain information on the kinetic processes at play. For investigating the electron acceleration in a large-scale phenomenon such as a flare, fully kinetic methods (e.g.\ Particle-in-Cell) are to date too computationally expensive, even when employing reduced scale separation (such as a reduced proton-to-electron mass ratio; see e.g.\ \citealt{Baumann2013}). Therefore, fully kinetic simulations are only employed to study very limited domain sizes (e.g.\ \citealt{guo2014apj,Li2019}); for larger simulations, test-particle methods can instead be employed owing to their relatively reduced costs. Test particles evolved on top of an MHD simulation do not provide feedback to the MHD fields, lacking self-consistent kinetic mechanisms. Nevertheless, test particles can be used to qualitatively study particle motion and acceleration and in simulations of coronal flares,  even as large as the entire flare loop {(e.g.\ \citealt{woodneukirch2005,gordovskyy2010,Gordovskyy2014,threlfall2015,zhou2015,zhou2016,Gordovskyy2020,Kong2022})}. The same approach can also adopt idealized setups such as coalescing magnetic islands, where extreme resolutions in MHD (obtained via grid adaptivity) can help identify sites of particle trapping (e.g.\ \citealt{zhao2021}).

{In this work, we utilize test-particle simulations to investigate the particle acceleration scenario suggested by \citet{Fang2016}, i.e.\ the turbulent acceleration of electrons inside the looptop. The test particle evolve in an MHD background for which we employ a simplified model of chromospheric evaporation in the post-flare loop state. This model is somewhat agnostic of the primary reconnection process above the looptop, which we do not model. As such, we simply assume that footpoint heating occurs and fluid flows meet and mix in the looptop region. Our objectives are to determine whether KHI turbulence in this region can confine electrons over long times inside the looptop, and whether these electrons can also experience acceleration to high energies to produce the strong looptop HXR sources that have been observed. To do so, we employ state-of-the-art MHD simulations of looptop turbulence first presented in \citet{Ruan2018} and analyze the trapping and acceleration of a population of test {electrons} inside the looptop. We then calculate whether the HXR emission obtained in these runs is compatible with observations.}

This paper is organized as follows: in Sec.~\ref{sec:model} we review the MHD and test-particle models employed in our looptop simulations. In Sec.~\ref{sec:trapping} we investigate the mechanism of particle trapping inside the MHD-simulated coronal looptop. In Sec.~\ref{sec:energization} we quantify particle acceleration in our test-particle runs. In Sec.~\ref{sec:hxr} we discuss whether the measured acceleration is compatible with HXR emission in looptops. Finally, in Sec.~\ref{sec:conclusions} we summarize our results.

\begin{figure*}
\centering
\includegraphics[width=1\textwidth, trim={0mm 10mm 0mm 0mm}, clip]{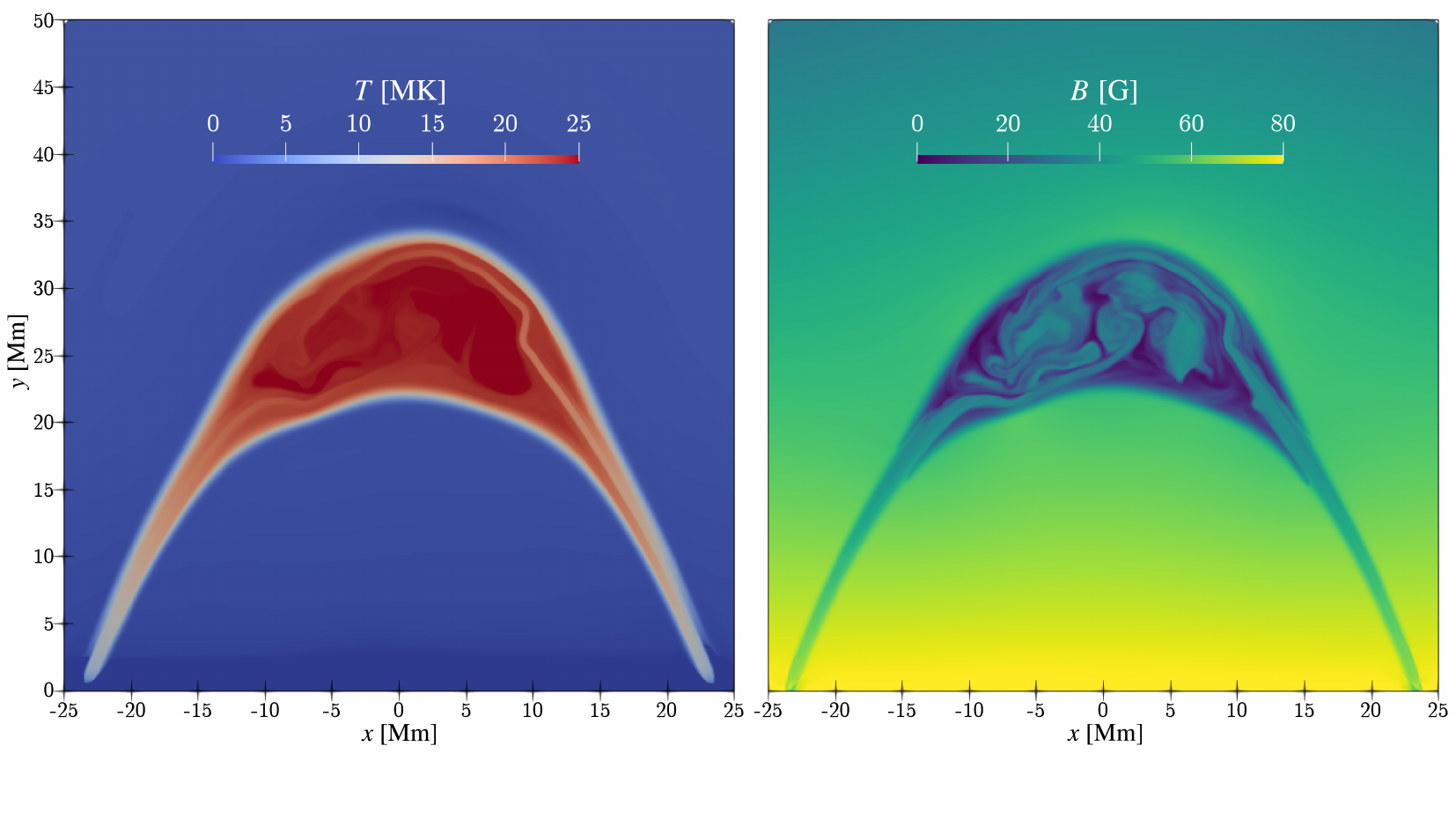}
\caption{Snapshot at $t=2.5t_0=186~\mathrm{s}$ of one of the MHD simulations of coronal loops (case C) used as initial conditions for the test-particle runs. Left: Spatial distribution of the temperature. Right: Spatial distribution of the magnetic-field strength.}
\label{fig:setup}
\end{figure*}

\section{Test-particle model in ideal MHD}
\label{sec:model}

\subsection{MHD Model for a Coronal Loop}
\label{sec:model_mhd}

Our MHD simulations are performed with the open-source \texttt{MPI-AMRVAC} code (\citealt{xia2018,keppens2023}). We run three ideal-MHD simulations (i.e.\ without including resistivity or other diffusion terms) with a simplified, two-dimensional (2.5D, i.e.\ including all three vector components) flare model, in which the reconnection current sheet is not included. The full setup is explained in detail in \cite{Ruan2018}, and we briefly summarize it here. The corona and the chromosphere are included at the bottom boundary of our setup, for which we employ a numerical domain of $-40~\mathrm{Mm} \leq x \leq 40~\mathrm{Mm}$ and $0~\mathrm{Mm} \leq y \leq 50~\mathrm{Mm}$. The base numerical resolution consists of $128 \times 80$ cells; via adaptive mesh refinement (AMR) we attain an effective resolution of $2048 \times 1280$ cells at the highest AMR level (using 5 levels). This implies an effective resolution of about 40~km, rivaling the observational limits. A potential magnetic field is set up in this configuration, which includes several magnetic arcades in the atmosphere. Localized heating is added at the chromospheric footpoints of selected magnetic arcades to produce evaporation flows. These flows generate a flare loop by filling the magnetic arcades with hot, dense plasma. Before the evaporation occurs, the magnetic-field strength at the footpoints is 80~G, and at the looptop 50~G.

The three runs differ by the amount of heat injected at the footpoints, from which fast plasma flows rise to then meet at the looptop. Where the streams meet, the KHI can develop and cause a turbulent cascade. The footpoint heating rate is given by Eqs.~(12-16) of \cite{Ruan2018}; in particular, we vary the heat injection via the parameter $c_1$ in their Eq.~(14). The resulting three simulations have $c_1 = 4.29 \times 10^{12}$, $8.24 \times 10^{12}$, $1.288 \times 10^{13} ~ \mathrm{erg~cm^{-2}}$, which we label as cases A, B, and C respectively. The parameter $\lambda_t$ in Eq.~(15) of \cite{Ruan2018} is set to $60$~s in all three cases, and other parameters are provided in Sec.~2 of their paper. These parameters are inspired by energy estimates from observations: assuming that the depth of the loop in the out-of-plane direction is $\sim10~\mathrm{Mm}$, then the energy injected into the chromosphere is of order $\sim10^{30}~\mathrm{erg}$, within the range of M-class flares (e.g.\ \citealt{Aschwanden2016}). {These parameters also provide maximum energy deposition rates of around $3\times10^{10}~\mathrm{erg/cm}^2\mathrm{/s}$ at the footpoints, which align well with observations and modeling estimates (e.g.\ \citealt{allred2015} and references therein).}

Due to the different amounts of heat injected, the three MHD runs evolve in a substantially different way, with case C developing clear turbulent structures that, over time, cascade towards smaller length scales. We let all cases evolve until $t=2.5t_0$, where the unit time in this work is $t_0=86~\mathrm{s}$. {This unit time is the time scale for acoustic waves to travel a distance $L_0 = 10~\mathrm{Mm}$ in a typical coronal-plasma environment with temperature $T_0 = 1~\mathrm{MK}$.} The spatial distribution of temperature and magnetic-field strength at $t=2.5t_0$ is shown in Fig.~\ref{fig:setup} for case C, i.e.\ the case with the strongest heat injection at the footpoints. We can observe that the upper part of the looptop is characterized by high temperatures ($\gtrsim20~\mathrm{MK}$); this region contains turbulent magnetic-field structures, and the looptop plasma is confined inside the loop by the much stronger ambient magnetic field outside.

\begin{figure*}
\centering
\includegraphics[width=1\textwidth, trim={0mm 0mm 85mm 0mm}, clip]{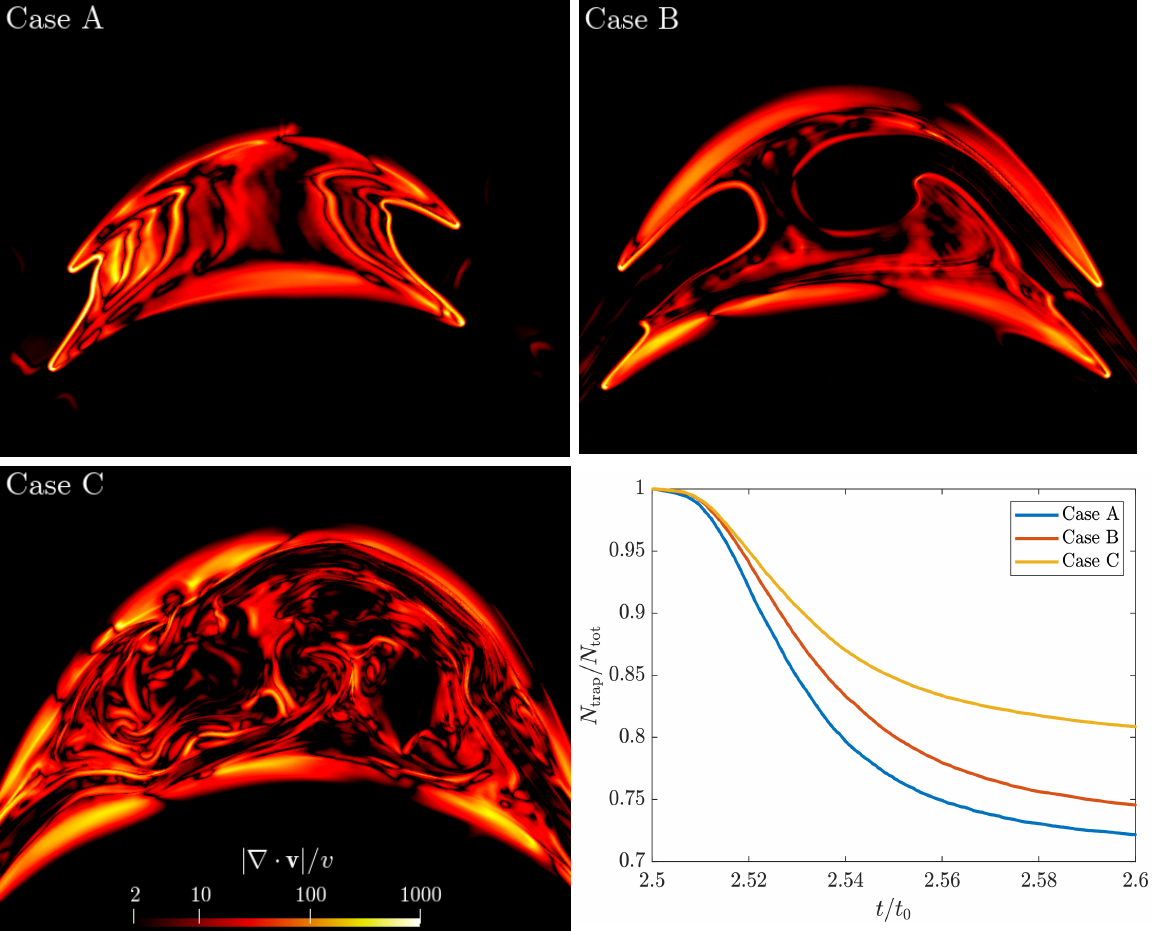}
\caption{Spatial distribution of $|\grad\bcdot\vecv|/v$ inside the looptop at $t=2.5t_0$ for the three MHD simulations considered. As the level of turbulence increases due to stronger heat injection from the footpoints (from case A to C, see Sec.~\ref{sec:model}), the looptop region features progressively more numerous turbulent structures, where the fluid velocity is characterized by sharp gradients. The bottom-right panel shows the fraction of {electrons} (with respect to the initial number) trapped inside the looptop over time.}
\label{fig:trap}
\end{figure*}

\subsection{Test Particles: Guiding-center Approximation}

To study the dynamics of electrons in the looptop turbulence of our MHD simulations, we evolve ensembles of {electrons} using the test-particle module of \texttt{MPI-AMRVAC 3.0} (\citealt{keppens2023}). These particle ensembles are tracked in the time-evolving MHD background described in Sec.~\ref{sec:model_mhd}, starting from the MHD state at $t=2.5t_0$. {Electrons} are initialized according to a Maxwellian distribution with temperature $T=20~\mathrm{MK}$ (approximately the average temperature inside the looptop, see Sec.~\ref{sec:trapping}); since the average magnetic-field strength inside the looptop is $B\sim 5\mbox{--}40~\mathrm{G}$ (see Fig.~\ref{fig:setup}), the maximum electron gyroradius $\rho_\mathrm{C} = \sqrt{m_ekT}/(|q_e|B) \sim 1~\mathrm{m}$ (where $m_e$ and $q_e$ are the electron mass and charge) at initialization. Because our MHD system size is of order $10^7~\mathrm{m}$, spatially resolving the gyromotion on the numerical grid would require unachievable resolution; the same applies for the gyration time scales, which are much faster than the MHD dynamical time. For this reason, we choose to evolve our test particles according to the guiding-center equations of motion, which are appropriate when the gyroradius has negligible size and the gyrofrequency is very large.

In our simulations we thus solve the relativistic equations of motion under the guiding-center approximation (GCA), in which the particle gyration around magnetic-field lines is averaged over, and only the motion of the particle guiding center is considered (e.g.\ \citealt{vandervoort1960,northrop1963}). In this paradigm, the spatial part of the particle four-velocity $\vecu=\vecv\gamma$ (with the Lorentz factor $\gamma=1/\sqrt{1-v^2/c^2}=\sqrt{1+u^2/c^2}$, where $c$ is the speed of light) is split into the parallel and perpendicular\footnote{Note that with $\uperp$ we indicate the perpendicular particle velocity \emph{linked to the particle's gyromotion}, i.e.\ excluding the velocity drift terms that determine the guiding-center motion across magnetic-field lines. These terms are typically much smaller than $\upar$, hence we can safely approximate $\uperp^2\simeq{u^2-\upar^2}$. If needed, a better approximation is given by $\uperp^2\simeq{u^2-\upar^2-v_E^2\gamma^2}$, since $\vecv_E$ is the dominant drift term (see e.g.\ \citealt{bacchini2020}).} components $\upar=\vecu\bcdot\vecb$ and $\uperp\simeq\sqrt{u^2-\upar^2}$ with respect to the magnetic field $\vecB$ with unit vector $\vecb=\vecB/B$. For a charged particle of mass $m$ and charge $q$, the guiding-center position $\vecR$, parallel four-velocity $\upar$, and magnetic moment $\mu=m\uperp^2/(2B\kappa)$ evolve as
\begin{equation}
  \frac{\rmd\vecR}{\rmd t} = \frac{\upar}{\gamma}\vecb + \vecv_E +\vecvcurv +\vecvpol +\vecvgradB +\vecvrel,
  \label{eq:gcaR}
\end{equation}
\begin{equation}
  \frac{\rmd\upar}{\rmd t} = \frac{q}{m}\Epar + \acurv + \agradB,
  \label{eq:gcaupar}
\end{equation}
\begin{equation}
 \frac{\rmd\mu}{\rmd t}=0,
 \label{eq:gcamu}
\end{equation}
where $\Epar=\vecE\bcdot\vecb$. Here, the motion of the particle's guiding center (eq.~\eqref{eq:gcaR}) is described as a superposition of motions along and across magnetic-field lines, indicated by a number of ``drift'' velocity terms. The dominant term $\vecv_E=\vecE\btimes\vecB/B^2$ is the ``$\vecE\btimes\vecB$'' drift, with associated Lorentz factor $\kappa=1/\sqrt{1-v_E^2/c^2}$. In addition to $\vecv_E$, the other drift terms are the curvature drift,
\begin{equation}
\vecvcurv = \frac{mc\kappa^2}{qB} \vecb\btimes \left[ \frac{\upar^2}{\gamma}\left(\vecb\bcdot\grad\right)\vecb+\upar\left(\vecv_E\bcdot\grad\right)\vecb\right],
\end{equation}
the polarization drift,
\begin{equation}
\vecvpol=\frac{mc\kappa^2}{qB} \vecb\btimes \left[ \upar\left(\vecb\bcdot\grad\right)\vecv_E+\gamma\left(\vecv_E\bcdot\grad\right)\vecv_E\right],
\end{equation}
the mirror (or ``$\grad B$'') drift,
\begin{equation}
\vecvgradB = \frac{\mu c\kappa^2}{\gamma qB} \vecb\btimes\grad \left( B/\kappa \right),
\end{equation}
and the relativistic drift,
\begin{equation}
\vecvrel =\frac{\upar \Epar\kappa^2}{c\gamma B}\vecb\btimes \vecv_E.
\end{equation}
Likewise, the parallel momentum evolves (eq.~\eqref{eq:gcaupar}) according to the parallel-acceleration term $\apar=q\Epar/m$, as well as the curvature acceleration,
\begin{equation}
\acurv = \vecv_E\bcdot\left[ \upar\left(\vecb\bcdot\grad\right)\vecb+\gamma\left(\vecv_E\bcdot\grad\right)\vecb\right],
\end{equation}
and the mirror acceleration term,
\begin{equation}
\agradB = -\frac{\mu}{m\gamma} \vecb\bcdot \grad\left( B/\kappa \right).
\end{equation}
Our Eq.~(\ref{eq:gcamu}) adopts the usual adiabatic-invariant assumption, such that the magnetic moment remains constant. {The use of GCA equations in test-particle simulations is customary in solar contexts (e.g.\ \citealt{woodneukirch2005,gordovskyy2010,threlfall2015})}; in all these expressions, we ignored terms proportional to time derivatives of the electromagnetic fields, under the assumption that particle dynamics takes place on much faster time scales than those over which MHD fields typically evolve (see e.g.\ \citealt{ripperda2017a,ripperda2017b,ripperda2018a}). Eqs.~\eqref{eq:gcaR}--\eqref{eq:gcamu} are solved in \texttt{MPI-AMRVAC} with a fourth-order Runge-Kutta method with adaptive stepsize (see Sec.~6.1 of \citealt{keppens2023}).



\section{Particle Trapping in MHD Looptop Turbulence}
\label{sec:trapping}

First, we wish to verify whether turbulence in the MHD looptop can promote particle trapping, as conjectured by \cite{Fang2016}. We thus consider the three MHD simulations (cases A, B, and C) of an SXR coronal loop introduced in Sec.~\ref{sec:model}. As a qualitative measure of the presence of turbulent structures inside the looptop, in Fig.~\ref{fig:trap} we show for all cases the spatial distribution of $|\grad\bcdot\vecv/v|$ in the looptop region at $t=2.5t_0$ (i.e.\ the end time of the preliminary MHD runs). We observe that case C (bottom-left panel) presents numerous turbulent structures with strong and chaotically distributed velocity gradients. Conversely, in cases A and B (top-left and top-right panels), the turbulence is not well developed and the looptop region features more coherent, larger-scale plasma flows.

\begin{figure*}
\centering
\includegraphics[width=1\textwidth, trim={0mm 0mm 90mm 0mm}, clip]{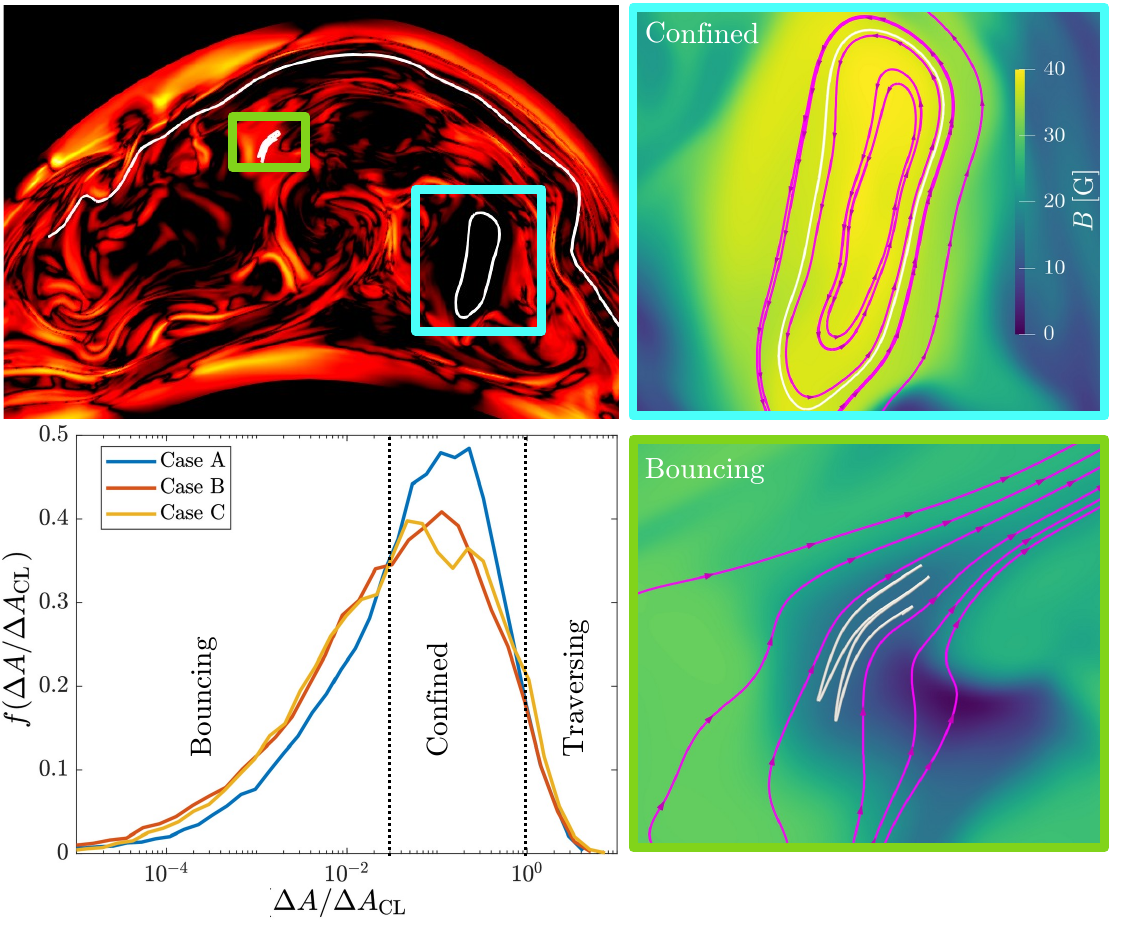}
\caption{Different types of electron trajectories inside a coronal looptop. Top-left panel: representative ``traversing'' (i.e.\ crossing the entire looptop), ``confined'', and ``bouncing'' trajectories (white lines). Top-right panel: zoomed-in view of a confined trajectory inside a large magnetic island (with magnetic-field lines shown in purple). Bottom-right panel: zoomed-in view of a bouncing trajectory between magnetic islands. Bottom-left panel: distribution of sizes of the spatial regions spanned by particle trajectories in the three different MHD simulations. As the level of turbulence increases (from case A to C), trajectories spanning smaller areas become more numerous. {Approximate thresholds on the size of the area spanned by different types of trajectories are indicated with dashed vertical lines.}}
\label{fig:paths}
\end{figure*}

At $t=2.5t_0$, we inject $10^6$ test electrons in each of the three MHD simulations. Particles are generated at random positions inside the looptop\footnote{A particle is considered ``inside'' the looptop if the MHD temperature at the particle position is above a specific threshold $T_\mathrm{min}=6.5~\mathrm{MK}$. This is approximately the temperature at the looptop edge, before a sharp decrease to the much lower ambient temperature (see Fig.~\ref{fig:setup}).} and with velocities drawn from an isotropic Maxwellian distribution with temperature $T=20~\mathrm{MK}$ (approximately the average temperature inside the looptop at $t=2.5t_0$). We evolve this ensemble of {electrons} according to the GCA eqs.~\eqref{eq:gcaR}--\eqref{eq:gcamu} until $t=2.6t_0$. During this time, we progressively delete particles that leave the looptop region, and keep track of the number of particles that remain inside that region. The evolution in time of the fraction of trapped {electrons} over the total {electrons} initially injected is shown in the bottom-right panel of Fig.~\ref{fig:trap}. Over time, particles progressively leave the looptop and the fraction of trapped particles decreases. The rate of particle escape starts slowing down in all cases around $t=2.55t_0$, and around $t=2.58t_0$ the fraction of remaining particles is stabilizing. For cases A and B, where turbulence is weak, the fraction of trapped {electrons} at the end of the integration time is comparatively (up to 10\%) smaller than for case C, i.e.\ the case where turbulence is better developed.

To understand the difference in trapping efficiency for the three cases, we analyze single-particle trajectories. The overall behavior of {electron} ensembles is similar between the three MHD simulations: the majority ($>70\%$) of particles are trapped inside the looptop .for long times; {electrons} generally cannot escape the looptop from the top or bottom interfaces, where the magnetic-field strength increases steeply towards the looptop exterior and therefore acts as an efficient, large-scale magnetic mirror. {Electrons} are instead typically observed leaving the looptop from the lower regions, close to the footpoints. In all cases, we observe that particle trajectories can be divided into three main classes: i) ``Traversing'' trajectories that cross the entire loop structure, eventually leaving the looptop in the vicinity of the footpoinrs; ii) ``Confined'' trajectories, where {electrons} travel along closed magnetic-field lines inside magnetic islands; and iii) ``Bouncing'' trajectories, where {electrons} follow a rapid, oscillating motion between magnetic islands. Representative trajectories (corresponding to a time period $t\in[2.5,2.6]t_0$, where the MHD background is largely unchanged) for the three types of trajectories are shown for case C in Fig.~\ref{fig:paths} (top-left panel).

Each class of trajectories can be qualitatively described as follows:
\begin{itemize}
    \item \emph{Traversing} particles follow long paths, exploring the looptop from one end to the other (white line in top-left panel in Fig.~\ref{fig:paths}). These trajectories exhibit increasing distortion as the level of turbulence increases: for cases A and B (underdeveloped turbulence), traversing particles closely follow unperturbed magnetic-field lines, while for case C (well-developed turbulence), clear scattering patterns emerge in the trajectories, due to particles encountering many regions of alternating magnetic fields. Even though all traversing particles eventually leave the looptop, due to these scattering effects the length of traversing trajectories increases with the level of turbulence. As a consequence, the time taken for traversing particles to leave the looptop also increases proportionally, producing higher trapping efficiencies.

    \item \emph{Confined} particles follow approximately closed (in the $xy$-plane) trajectories inside the large magnetic islands belonging to the looptop (top-right panel in Fig.~\ref{fig:paths}). The motion of these particles in the $xy$-plane is predominantly parallel to the in-plane magnetic-field lines forming these islands, without strong scattering. Particles traveling along these trajectories may remain confined for long times, leaving magnetic islands only when the magnetic-field structure in the confining region changes significantly during the MHD evolution.

    \item \emph{Bouncing} particles exhibit fast oscillatory motion in limited regions of space between large magnetic islands (bottom-right panel in Fig.~\ref{fig:paths}). These particles are reflected at the island interfaces, where the local steep increase in the magnetic-field strength acts as an efficient mirror.
\end{itemize}

To qualitatively relate the presence of different particle trajectories with the level of development of turbulence inside the looptop, we adopt the following strategy: for each particle, we measure the ``area'' $\Delta A$ spanned by its trajectory, by simply approximating each area as a rectangle bounded by the maximum and minimum coordinate reached by a particle in each direction. We normalize the measured areas by the area $\Delta A_\mathrm{CL}\simeq350~(\mathrm{Mm})^2$ of the upper looptop region, to qualitatively compare the spatial size of the region explored by {electrons} with the size of the whole turbulent looptop. Then, we construct the distribution function $f(\Delta A/\Delta A_\mathrm{CL})=\rmd N/\rmd(\Delta A/\Delta A_\mathrm{CL})$ of the range of areas spanned by all {electrons} in the three MHD simulations. This is shown in Fig.~\ref{fig:paths} (bottom-left panel), where we measure a clear increase in the number of {electrons} spanning smaller areas when turbulence is more developed (e.g.\ case C). Conversely, more {electrons} spanning larger areas are present when turbulence is less developed (e.g.\ case A). This indicates that MHD turbulence inside coronal looptops can achieve a high trapping efficiency by confining {electrons} in smaller regions of space, inside or between magnetic islands formed by the turbulence dynamics. In particular, we measure a net increase in the total number of bouncing trajectories, which are much more numerous for case C, where turbulence is well developed and more magnetic islands are present. In Fig.~\ref{fig:paths} (bottom-left panel), we have subdivided the range of $\Delta A/\Delta A_\mathrm{CL}$ {with approximate thresholds} to indicate traversing trajectories (with areas comparable with the looptop size, $\Delta A/\Delta A_\mathrm{CL}>1$), confined trajectories (exploring a significant fraction of the looptop region, $0.03<\Delta A/\Delta A_\mathrm{CL}<1$), and bouncing trajectories (with areas much smaller than the size of the looptop, $\Delta A/\Delta A_\mathrm{CL}<0.03$). {We emphasize that these thresholds are only indicative, and mainly serve to guide the eye when qualitatively classifying different trajectory types. We have however verified that, below the indicated threshold, no confined trajectories exist, and particles are only found on bouncing orbits.}

\begin{figure}
\centering
\includegraphics[width=0.96\textwidth, trim={0mm 0mm 140mm 0mm}, clip]{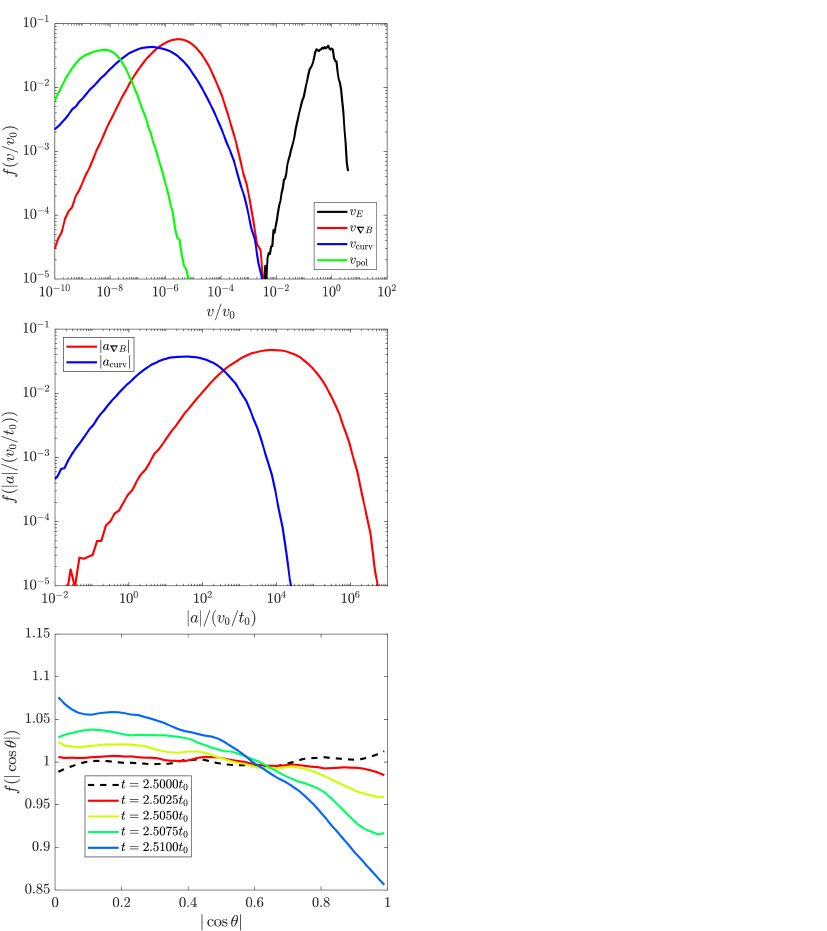}
\caption{Top and middle panels: Distribution at $t=2.5t_0$ (the time of the initial particle injection) of the magnitude of drift-velocity and parallel-acceleration terms, {respectively}, measured inside the looptop for case C, according to the GCA eqs.~\eqref{eq:gcaR}--\eqref{eq:gcamu} (without $\Epar$-related terms). The dominant drifts are due to $\vecE\btimes\vecB$ and mirror motion, and the dominant force term is related to mirror effects. Bottom panel: Distribution of pitch angles for $t\in[2.5,2.51]t_0$, showing that {electrons} experience rapid scattering across magnetic-field lines.}
\label{fig:drift_acc_pitch}
\end{figure}

The particle-trapping dynamics discussed above can be understood by considering the relative importance of drift and force terms, determined by the MHD background conditions, that appear on the right-hand side of eqs.~\eqref{eq:gcaR}--\eqref{eq:gcamu}. In Fig.~\ref{fig:drift_acc_pitch}, we show the initial (at $t=2.5t_0$, the time of the initial particle injection) distribution of the magnitude of all drift-velocity terms (top panel) and parallel-acceleration terms (middle panel) from the GCA equations\footnote{Note that we exclude terms related to $\Epar$, which vanishes in our ideal-MHD simulations.}, normalized by a reference fluid velocity $v_0\simeq1.16\times10^5~\mathrm{m/s}$. These terms determine the particle motion in our simulations; as expected from the ordering of drift terms in the governing equations, we observe that the dominant drift velocity is $\vecv_E$. Mirror and curvature drifts $\vecvgradB$ and $\vecvcurv$ are the second and third most important terms, with the polarization drift $\vecvpol$ contributing the least to particle motion. The distribution of parallel-acceleration terms in Fig.~\ref{fig:drift_acc_pitch}, in accordance with that of drift-velocity terms, shows that mirror acceleration $\agradB$ dominates over curvature acceleration $\acurv$.

\begin{figure*}
\centering
\includegraphics[width=0.9\textwidth, trim={0mm 0mm 30mm 0mm}, clip]{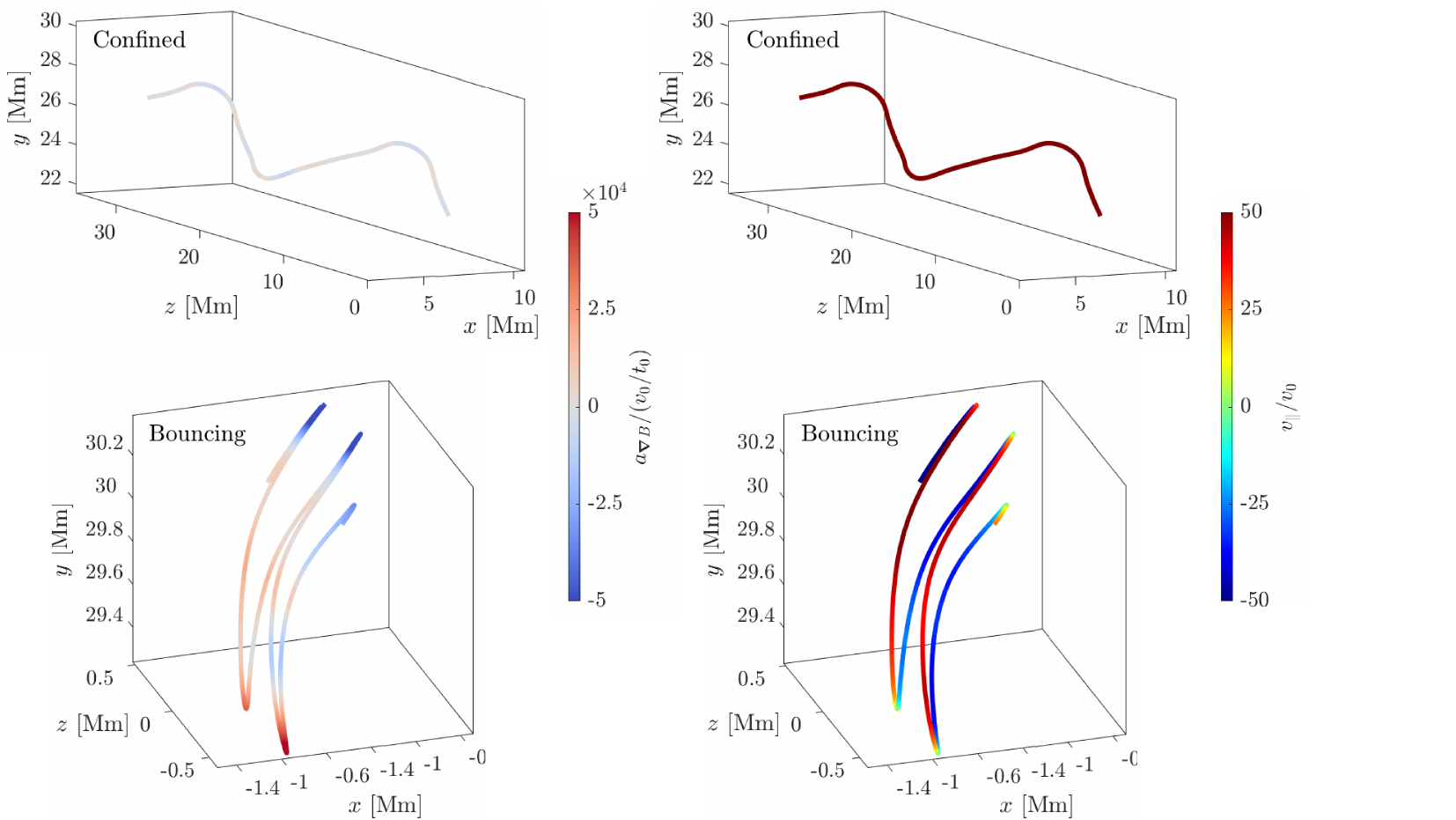}
\caption{A portion of the same representative ``confined'' (top row) and ``bouncing'' (bottom row) trajectories of Fig.~\ref{fig:paths}, shown here in three dimensions and colored by $\agradB$ (left) and $v_\|=\upar/\gamma$ (right). While confined particles mostly follow magnetic-field lines, bouncing particles are continuously reflected by mirror forces.}
\label{fig:midpathshortpath}
\end{figure*}

Finally, in the bottom panel of Fig.~\ref{fig:drift_acc_pitch} we show the evolution of the distribution of $|\cos\theta|$ over $t\in[2.5,2.51]t_0$, where the pitch angle $\theta=\tan^{-1}(u_\perp/u_{\|})$. From an initial isotropic distribution at $t=2.5t_0$, we observe a gradual increase in the number of {electrons} with small $|\cos\theta|$ (i.e.\ large pitch angles), and a corresponding decrease in the number of {electrons} with large $|\cos\theta|$ (i.e.\ small pitch angles). This is a consequence of dominant mirror forces acting on the {electrons}: within a relatively short time ($\delta t=0.01t_0$), a large fraction of {electrons} experience strong scattering and start following orbits that cross magnetic-field lines. In a turbulent flow such as that created inside the looptop, this scattering (and associated mirror forces) is promoted by numerous structures of alternating magnetic field along random directions. Overall, {electrons} experiencing continuous scattering are much less likely to follow open field lines, and are therefore confined inside the looptop for longer times. These scattered {electrons} precisely correspond to those following ``bouncing'' trajectories as shown in Fig.~\ref{fig:paths}, while ``confined'' {electrons}, although trapped, simply follow closed field lines. Since these two types of trajectories correspond to the long-lived particle populations that remain inside the looptop over long times, we now analyze the dynamics of these {electrons} more in detail.

\emph{Confined trajectories}: The distinguishing feature of these trajectories is the trapping effect that occurs \emph{inside} magnetic islands, i.e.\ regions of strong $B$ characterized by closed field lines in the $xy$-plane. Particles on confined trajectories tightly follow closed field lines in a predominantly parallel motion. In Fig.~\ref{fig:midpathshortpath} (top row), we show the same confined trajectory of Fig.~\ref{fig:paths}, now in three dimensions (including the out-of-plane motion); the trajectory is colored by the local $\agradB$ (left panel) and $v_{\|}=\upar/\gamma$ (right panel). In addition to forming closed loops in the $xy$-plane, confined particles also travel long distances along $z$ (the direction of the guide field) without experiencing strong mirror acceleration and while maintaining an approximately constant (in magnitude and sign) parallel velocity. The lack of strong scattering and acceleration determines a tight confinement inside the same magnetic island for long times.

\emph{Bouncing trajectories}: Bouncing particles are characterized by a rapid, oscillating motion trapped \emph{between} magnetic islands. In Fig.~\ref{fig:midpathshortpath} (bottom row) we show the same bouncing trajectory of Fig.~\ref{fig:paths}, here in three dimensions and colored by $\agradB$ (left panel) and $v_{\|}=\upar/\gamma$ (right panel). This particle only travels a short distance along $z$, instead moving predominantly in the $xy$-plane while bouncing between regions of large magnetic-field gradients. Strong mirror forces act on this particle, converting parallel motion into perpendicular motion, causing an inversion in the particle trajectory and a drift across magnetic-field lines.

\begin{figure*}
\centering
\includegraphics[width=1\textwidth, trim={0mm 0mm 0mm 0mm}, clip]{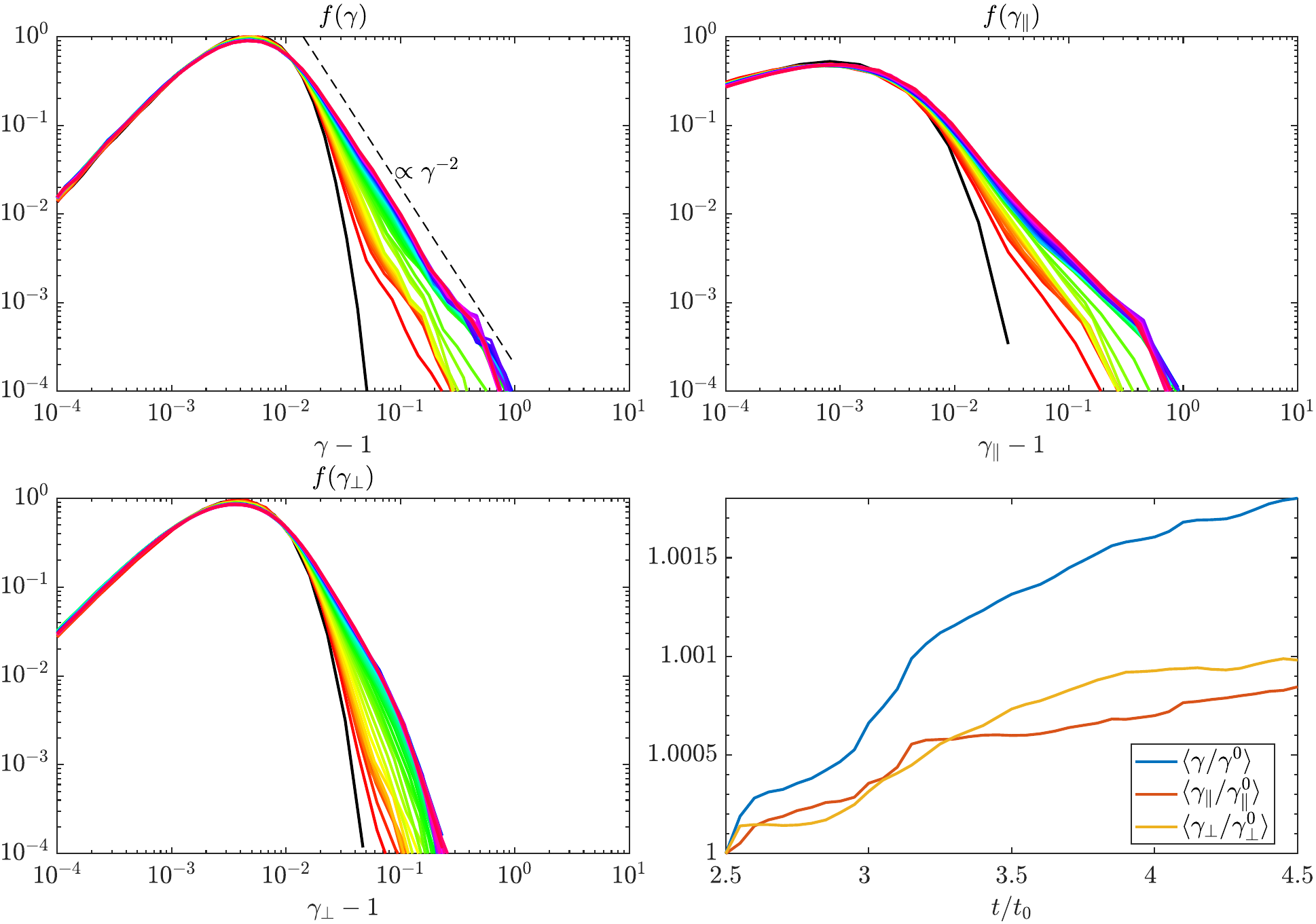}
\caption{Evolution of the total particle energy (top-left panel), parallel energy (top-right panel), and perpendicular energy (bottom-left panel) for $t\in[2.5,4.5]t_0$. Energy distributions are evenly spaced in time with a cadence $\delta t=0.05t_0$. The evolution of the mean energy relative to the initial energy, $\gamma/\gamma^0$, is shown in the bottom-right panel for the same time period.}
\label{fig:res_study_amr7_dist}
\end{figure*}

With this analysis, we are able to {identify the forces that determine qualitatively different particle trajectories (i.e.\ confined and bouncing), namely inside and between magnetic islands (i.e.\ structures of coherent fields). Confined particles are found in regions of strong field and are therefore tightly bound to magnetic-field lines. Conversely, bouncing particles are found in regions of weaker fields, and experience strong mirror forces and fast oscillatory motion when encountering island boundaries; this leads to scattering and conversion of parallel momentum into perpendicular momentum and vice versa.}

In the next Sections, we analyze the energetics of looptop-trapped particles more in detail.

\section{Particle Energization in MHD Looptop Turbulence}
\label{sec:energization}

{Electrons} trapped inside the looptop for long times can experience energization; in the absence of parallel electric fields (as in our ideal-MHD simulations), the main forces acting on particles (therefore determining energization) are represented by curvature and mirror effects, as indicated in Fig.~\ref{fig:drift_acc_pitch} (middle panel). Inside the turbulent looptop, it is expected that these effects manifest via continuous scattering of {electrons} across turbulent magnetic structures, promoting Fermi-type acceleration mechanisms (e.g.\ \citealt{guo2019,zhangxiang2021,lemoine2022}). To investigate this phenomenon, here we analyze the time evolution of particle energy distributions in our simulations, taking care of distinguishing between parallel and perpendicular energization.

With the same MHD looptop setup employed in Sec.~\ref{sec:trapping}, we perform again test-particle simulations initializing 10$^6$ electrons in the looptop region. The initial distribution is again a Maxwellian with temperature $T=20~\mathrm{MK}$. Having assessed that stronger turbulence produces more efficient trapping (see previous Section), here we focus on the MHD run where turbulence is most developed (case C presented earlier). We let our particles evolve in the MHD background, starting from the state shown in Fig.~\ref{fig:setup} at $t=2.5t_0$ and running the simulation until $t=4.5t_0$, i.e.\ for two full MHD dynamical times. Because the time integration occurs over MHD time scales, we also concurrently evolve the MHD background; in addition, we allow the resolution to increase with respect to the initial state (which has a maximum AMR level of 5), by setting the maximum AMR level to 7. As a result, inside the turbulent looptop we now achieve an effective resolution of $8192 \times 5120$ cells, i.e.\ our minimum grid spacing is of order $\sim10~\mathrm{km}$. Note that this is still 4 orders of magnitude larger than the typical electron gyroradius, justifying the choice of the GCA paradigm.

In Fig.~\ref{fig:res_study_amr7_dist} we show the time evolution of particle energy distributions $f(\gamma)=\rmd N/\rmd\gamma$. The distributions are plotted with cadence $\delta t=0.05t_0$ for $t\in[2.5,4.5]t_0$. From the initial Maxwellian (shown in black), we observe progressive energization of {electrons} toward the high-energy range (top-left panel). The distribution develops a suprathermal tail with slope $\propto\gamma^{-2}$ and an upper cutoff around $\gamma\simeq2$, i.e.\ mildly relativistic energies are achieved in the particle population. In the top-right and bottom-left panels of Fig.~\ref{fig:res_study_amr7_dist} we also show the evolution of energy distributions in the parallel and perpendicular directions, expressed by $f(\gamma_\|) := f(\sqrt{1+\upar^2/c^2})$ and $f(\gamma_\perp) := f(\sqrt{1+\uperp^2/c^2})$ respectively. Here, we can observe that the main nonthermal features arise in the parallel energy, where a high-energy tail of constant slope develops. In the perpendicular energy, instead, suprathermal {electrons} are present but do not populate a well-defined tail with a constant characteristic slope. Finally, in the bottom-right panel we show the evolution in time of the average (over all particles) increase in total, parallel, and perpendicular energy with respect to the initial (at $t=2.5t_0$) energy of each particle, $\gamma^0$. We observe that, on average, {electrons} in the initial population have experienced an increase in $\gamma$ of $\gtrsim 0.1\%$; this increase in energy is, in our ideal-MHD simulation, entirely attributed to Fermi-like processes (see e.g.\ \citealt{lemoine2022}). Furthermore, the relative energization in the perpendicular direction is, on average, slightly stronger than in the parallel direction.
Considering that, as discussed in the previous Sections, many particles experience long-term confinement between magnetic islands, this suggests that the mirror effects at play are efficiently pumping energy into perpendicular particle motion.

In Fig.~\ref{fig:res_study_amr7_dist}, we also observe that the acceleration proceeds in stages. Between $t=2.5t_0$ and $t=2.55t_0$ (i.e.\ between the initial condition, in black, and the first plotted line, in red), {electrons} in the high-end of the energy spectrum are already rapidly pushed to form a mild suprathermal population. After this phase, the energization becomes progressively slower; a nonthermal tail arises and settles into a $\propto\gamma^{-2}$ slope and between $t=3.5t_0$ and $t=4.5t_0$ we do not measure significant changes in the energy distribution. This indicates that one MHD dynamical time ($\sim86$~s) suffices to produce an asymptotic distribution that does not significantly evolve over longer times, although on average {electrons} are still gaining energy at a low rate at the end of the run (as shown in the bottom-right panel) of Fig.~\ref{fig:res_study_amr7_dist}.

\begin{figure}
\centering
\includegraphics[width=.99\columnwidth, trim={0mm 0mm 0mm 0mm}, clip]{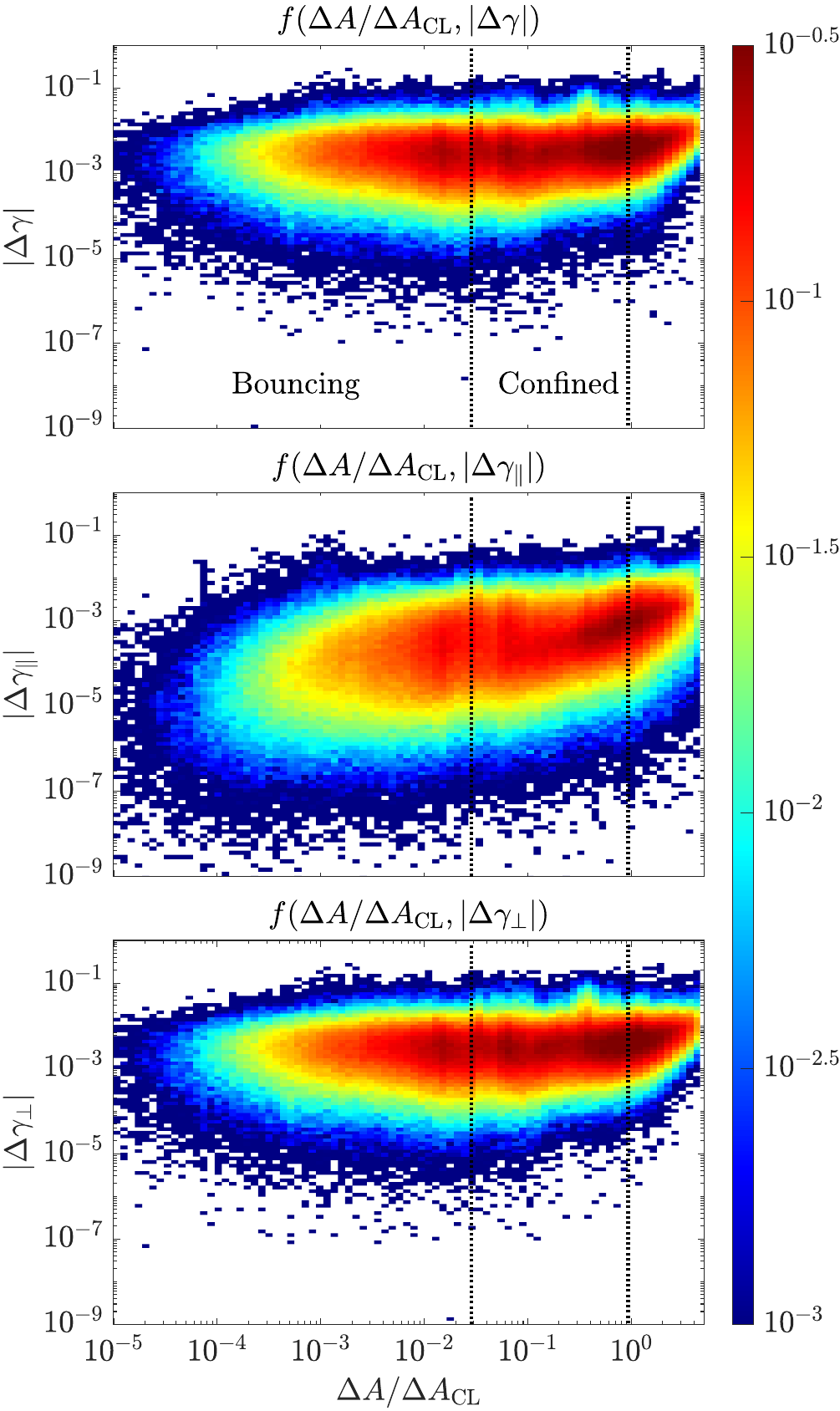}
\caption{Two-dimensional distribution of total (top panel), parallel (middle panel), and perpendicular (bottom panel) energy gain versus area spanned for all trajectories over the time interval $t\in[3.10, 3.15]t_0$.}
\label{fig:dAvsdg}
\end{figure}

Finally, we analyze the relation between particle trajectories and energy gain. In Fig.~\ref{fig:dAvsdg} we show the two-dimensional distribution of energy gain $\Delta\gamma=\gamma-\gamma^0$ versus the area spanned by particle trajectories $\Delta A/\Delta A_\mathrm{CL}$ (the latter defined as in the previous Section). To compute these distributions, we focus on the time interval $t\in[3.10, 3.15]t_0$ (hence here $\gamma^0$ is measured at $t=3.10t_0$), i.e.\ around halfway, and for a fraction 1/40, of the total simulation time; we also distinguish between total (top panel), parallel (middle panel), and perpendicular (bottom panel) energy gain. In each panel, we indicate the range of $\Delta A/\Delta A_\mathrm{CL}$ corresponding to bouncing and confined trajectories (as in Fig.~\ref{fig:paths}).

A first point of interest is that parallel-energy gain is generally restricted to $|\Delta\gamma_\||\sim 10^{-3}$ and is most prominent only for confined trajectories in a narrow range $\Delta A/\Delta A_\mathrm{CL}\in[0.5,1]$. Conversely, perpendicular-energy gain is detected up to $|\Delta\gamma_\perp|\gtrsim 10^{-2}$ for a large interval $\Delta A/\Delta A_\mathrm{CL}\in[10^{-2},1]$ spanning both confined and bouncing trajectories\footnote{We also detect energy gain for $\Delta A/\Delta A_\mathrm{CL}>1$, corresponding to traversing particles, which however eventually leave the looptop and are therefore of no particular interest here.}. Assuming that the time interval we consider is representative of the typical particle energization in developed MHD looptop turbulence, this result implies that significant parallel-energy gain is only achieved by confined particles traveling along the largest closed orbits inside magnetic islands ($\Delta A/\Delta A_\mathrm{CL}\simeq 1$, i.e.\ those that explore a region comparable in size to the whole looptop); these are only a small fraction of the total particles. Perpendicular-energy gain, instead, occurs rather generally for all particles (both bouncing and confined) and is stronger than parallel-energy gain. From Fig.~\ref{fig:res_study_amr7_dist} we also know that, over long times, parallel-energy gain tends to form nonthermal tails but is on average slightly smaller than perpendicular-energy gain. We can therefore conclude that perpendicular energy is rather ubiquitously pumped from the MHD turbulence into the particle motion, irrespective of the type of trajectory that {electrons} follow; while significant parallel-energy gain only occurs for a small population of {electrons} confined on specific long-lived orbits without significant scattering.

\section{Production of hard X-rays from accelerated electrons}
\label{sec:hxr}

\begin{figure*}
\centering
\includegraphics[width=1\textwidth, trim={0mm 0mm 20mm 0mm}, clip]{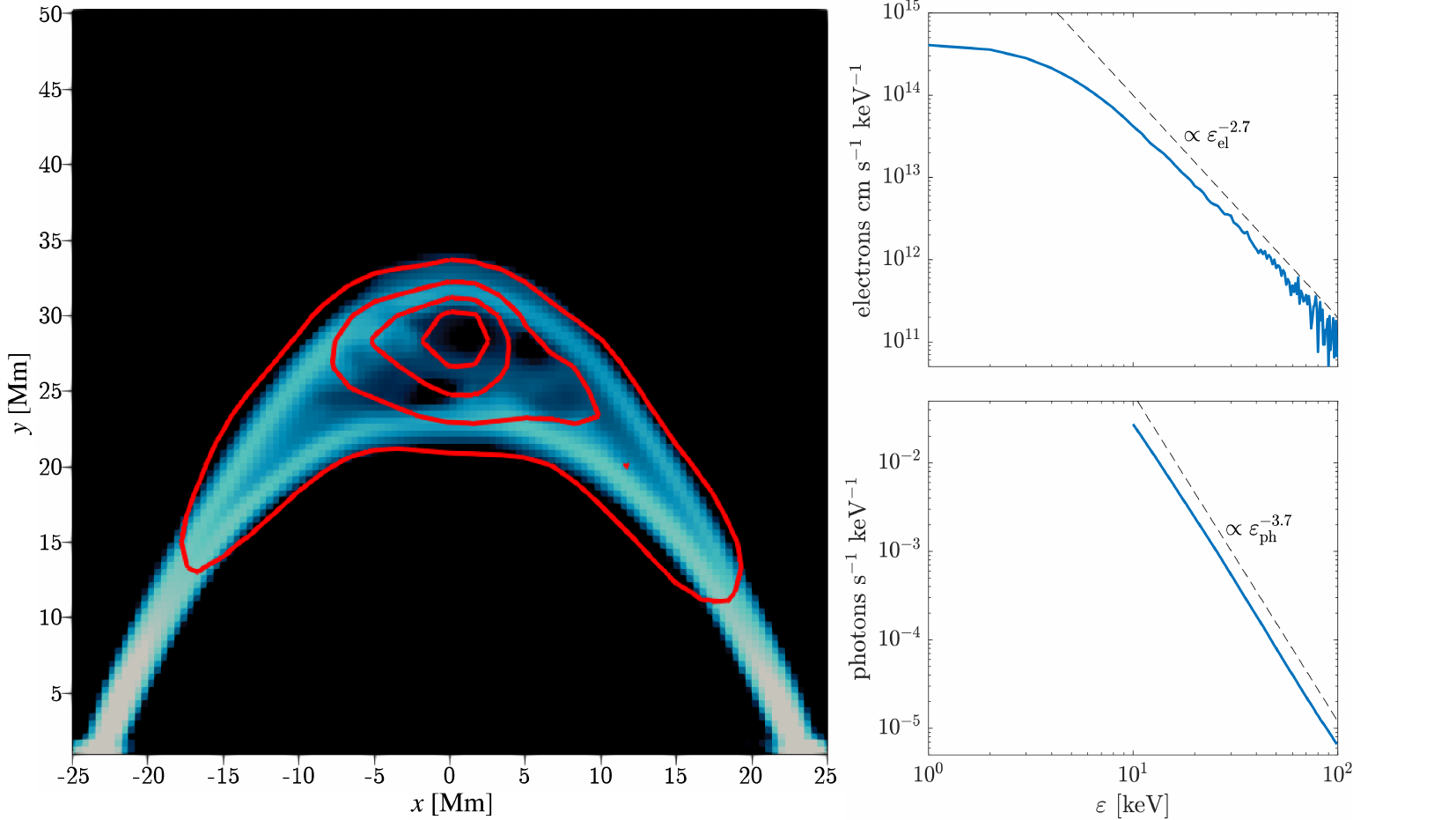}
\caption{Left panel: synthetic emission i) in the EUV at 131 {\AA} passband (in the background), computed from the MHD quantities at $t=4.5t_0$ with pixel size 0.6$''$ to mimic an SDO/AIA observation; and ii) in the HXR 10--20~keV energy range (as red isocontours of X-ray flux at 10\%, 30\%, 50\%, and 70\% of the maximum value), computed from the MHD and energetic electrons with pixel size 2.3$''$ to mimic a RHESSI observation. Right panels: spatially integrated test-particle flux spectrum (top) and derived HXR photon spectrum.}
\label{HXR}
\end{figure*}

We now turn our attention to the possibility of producing  HXRs from the plasma dynamics inside the looptop. In the scenario presented by \cite{Fang2016}, it is suggested that strong looptop HXR sources can be produced by energetic electrons by inverse Compton scattering of soft seed photons to higher energies. However, recent works have highlighted that inverse Compton scattering should be expected to have a much smaller contribution to looptop HXR emission than Bremsstrahlung, because of the low flux of SXR photons and the low collision rate between energetic electrons and the SXR photons (see e.g.\ \citealt{Ruan2018} and references therein). Here, we therefore consider HXR emission from  Bremsstrahlung as the dominant mechanism, and perform forward modeling of this emission using our particle data.

We synthesize HXR emission based on the spatial and energy distribution of the test {electrons} in the looptop. For ion-electron Bremsstrahlung-led HXR emission, the emissivity in terms of $\mathrm{photons\ s^{-1}\ cm^{-3}\ keV^{-1}}$ is given by
\begin{equation}
j(\varepsilon_\mathrm{ph}) = \int_{\varepsilon_\mathrm{ph}}^{\infty} n F(\varepsilon_\mathrm{el}) Q(\varepsilon_\mathrm{ph},\varepsilon_\mathrm{el}) \mathrm{d}\varepsilon_\mathrm{el},
\end{equation}
where $\varepsilon_\mathrm{ph}$ and $\varepsilon_\mathrm{el}$ are the photon and electron energy (in $\mathrm{keV}$), $n$ is local ion number density (in $\mathrm{cm}^{-3}$), $F$ is the energetic-electron flux (in $\mathrm{electrons\ s^{-1}\ cm^{-2}\ keV^{-1}}$), and $Q$ is the emission cross-section (in $\mathrm{cm^{2}}$) (see e.g.\ \citealp{Kontar2011}). To account for the contribution of our test {electrons} to the emissivity $j$ calculated at each spatial location $\bb{x}$, we compute for each $i$-th {electron}
\begin{equation}
j_i(\bb{x},\varepsilon_\mathrm{ph}) = n(\bb{x}) v_i Q(\varepsilon_\mathrm{ph},\varepsilon_{\mathrm{el},i}) \delta(|\bb{x}-\bb{x}_i|),
\end{equation}
where $\bb{x}_i$ is the location of the $i$-th test {electron}, $v_i$ is the {electron} speed, $\varepsilon_{\mathrm{el},i}$ is the {electron} energy, and $\delta$ is the Dirac delta function. The emission cross-section is taken from \cite{Haug1997}. To store the HXR flux, we employ an image mesh of pixel size $\sigma=2.3\arcsec$, equal to the pixel size of RHESSI observations (\citealt{Lin2002}). The brightness of the $k$-th pixel is therefore given by
\begin{equation}
\begin{aligned}
I_k  = \sum_{i} \iint \int_{\varepsilon^\mathrm{min}_{\mathrm{ph}}}^{\varepsilon^\mathrm{max}_{\mathrm{ph}}} j_i(\bb{x},\varepsilon_\mathrm{ph})
e^{\frac{-|\boldsymbol{x}-\boldsymbol{x}'|^2}{2 \sigma^2}} \mathrm{d} \varepsilon_\mathrm{ph} \mathrm{d}^2 \bb{x}\mathrm{d}^2 \bb{x}' \\
 = \sum_{i} \int \int_{\varepsilon^\mathrm{min}_{\mathrm{ph}}}^{\varepsilon^\mathrm{max}_{\mathrm{ph}}}
n(\bb{x}_i) v_i Q(\varepsilon_\mathrm{ph},\varepsilon_{\mathrm{el},i}) e^{\frac{-|\boldsymbol{x}_i-\boldsymbol{x}'|^2}{2 \sigma^2} } \mathrm{d} \varepsilon_\mathrm{ph} \mathrm{d}^2 \bb{x}',
\end{aligned}
\end{equation}
where multiplication by a Gaussian point-spread function mimics the instrument effect, with the pixel size $\sigma$ representing the variance.

The results of this calculation, carried out at $t=4.5t_0$ (i.e.\ the final time in our simulation), are shown in Fig.~\ref{HXR} (left panel), where we show isocontours of the HXR flux (in the 10--20~keV energy range) at 10\%, 30\%, 50\%, and 70\% of the maximum. The background color here represents synthetic EUV emission in the 131 {\AA} passband, calculated from the MHD quantities with a pixel size of $0.6''$ to mimic observations from SDO/AIA. In the same Figure, we show the spatially integrated test-particle flux spectrum (top right) and HXR photon spectrum (bottom right).
We observe that, due to the trapping of energetic electrons at coronal height, a strong coronal HXR source is produced. The coronal HXR source overlaps with the top of the bright loop in the EUV 131 {\AA} passband, similar to some observations (e.g.\ \citealt{Su2013}).
The HXR spectrum here has a single power-law distribution with a spectral index of $\sim$3.7. This is in accordance with observational data, where a spectral index around 4 is commonly measured in flare coronal HXR sources (e.g.\ \citealt{Petrosian2002,Battaglia2006,Gary2018}).

\section{Conclusions}
\label{sec:conclusions}
In this work, we presented a study of the dynamics of electrons in SXR, post-flare coronal loops. This is a first demonstration of the particle-tracing capabilities of \texttt{MPI-AMRVAC 3.0} (\citealt{keppens2023}), which we exploited to run 2.5D, ideal-MHD simulations of an entire post-flare loop in which turbulence driven by Kelvin--Helmholtz instabilities develops at the looptop. This was earlier identified as resulting from the observationally established flare-driven chromospheric evaporations that invade post-flare loops from the footpoint regions. The MHD setup is taken from \cite{Ruan2018}, considering three increasing rates of energy injection at the loop footpoints. {Our MHD simulations are agnostic of the main reconnection process above the looptop, which we do not model; rather, we assume footpoint heating (which comes e.g. from accelerated particle beams from the reconnection region, which sweep out laterally with the flare ribbons) driving fluid flows and creating post-flare looptop turbulence}. In this time-evolving, large-scale turbulence, we modeled ensembles ($\sim10^6$) of test electrons (which do not provide feedback to the MHD), characterizing their trapping and acceleration dynamics. Due to the vast difference between MHD and particle scales (e.g.\ in terms of the electron gyroradius), we resorted to the guiding-center approximation to evolve our test {electrons}.

First, we have studied the relation between the development of KHI turbulence and the trapping efficiency of the looptop region. We injected our test electrons in the looptop region, drawing their initial velocity from a Maxwellian distribution with temperature $T=20$~MK. By considering a short time interval (where the MHD background is practically unchanged), we have found that when turbulence is well-developed (in our case, due to fast flows rising from the loop footpoints), many more {electrons} can remain trapped inside the looptop. This is expected, since the turbulence provides an efficient scattering mechanism that causes particles to continuously bounce between alternating magnetic fields in the looptop, reducing the possibility of following open field lines that lead outside of the looptop region. By studying the types of particle trajectories we observe in the looptop turbulence, we indeed found that escaping (``traversing'') particles constitute a small population that follows long trajectories, crossing the whole turbulent region and leaving the looptop from the sides. Conversely, {electrons} trapped for long times either follow closed field lines inside magnetic islands (``confined'' trajectories) or rapidly  bounce between islands (``bouncing'' trajectories). We have shown that better-developed turbulence, in particular, corresponds to an increase in the number of confined and bouncing particles; their trapping is determined by strong mirror forces, which dominate (after $\vecE\times\vecB$ terms) among the guiding-center drifts. We have verified that the action of mirror terms, characteristic of developed turbulence, very rapidly scatters {electrons} to large pitch angles, determining drifts across magnetic-field lines that result in the bouncing motion of trapped particles.

We have then turned our attention to electron energization in MHD looptop turbulence. We initialized again our test {electrons} from a Maxwellian with $T=20$~MK, and evolved both the {electrons} and the MHD background for two MHD dynamical times (approximately $172$~s). In this way, we could track the evolution in time of particle energy distribution functions while the background concurrently evolves over multiple MHD timescales. In our ideal-MHD runs (where parallel electric fields are absent by definition), we expect Fermi processes to be the sole responsible of particle acceleration and to potentially create a power-law energy spectrum. We observed that, for the setup where turbulence is best developed, the initial Maxwellian distribution indeed develops a clear power-law tail that stabilizes to a characteristic $\propto\gamma^{-2}$ slope within one MHD time ($\sim 86$~s). This energy gain occurs in both the parallel and the perpendicular direction with respect to the local magnetic field, but the energization is qualitatively different: in the parallel direction, the energy distribution presents a power-law tail, which is instead absent in the perpendicular-energy distribution. We also measured the relative increase in energy (with respect to the initial value) per particle, finding that on average, {electrons} gain slightly more energy in the perpendicular direction, at least over long times. Finally, we have correlated the energy increase with the type of trajectory followed by each particle, finding that larger perpendicular-energy gain corresponds to bouncing trajectories confined within smaller regions of space. This indicates that the mirror terms previously discussed can efficiently act to energize trapped {electrons} in the perpendicular direction.

To build a link between our numerical experiments and possible observations, we have constructed synthetic radiation maps combining the MHD and test-particle data from our simulations. Assuming Bremsstrahlung-led HXR emission, we have calculated the photon fluxes that our energetic electrons can produce in the turbulent looptop MHD background, processing the data via an artificial pixel size of 2.6$''$ to mimic a RHESSI observation. We measured clear HXR radiation in the interior of the looptop in the range 10--20~keV, with an integrated HXR photon flux with characteristic slope $\propto\varepsilon_\mathrm{ph}^{-3.7}$. This is in good agreement with some observations (e.g.\ \citealt{Petrosian2002,Battaglia2006,Gary2018}); however, other observational studies report that coronal HXR sources can be located above the bright EUV SXR loop, suggesting that most of the electron acceleration could happen in the flare reconnection region (see e.g.\ \citealt{Masuda1994,chen2020NatAs}) instead of the looptop region that we consider here. In this context, it could be of interest to see whether high resolution HXR imagers, like the ASO-S/HXI instrument (\citealt{gan2019}), can reveal fine-structured, looptop HXR sources in flare systems, where potentially HXR emission may arise both from the region above the looptop (i.e.\ reconnection-outflow and termination-shock regions) and from underlying, post-flare turbulent looptops, as studied here.

The aim of this work was to build a first link between test-particle simulations, so far relatively underexplored in solar applications (e.g.\ \citealt{Kong2022} and references therein), and potential observables produced by energetic {electrons} originating in the turbulent coronal plasma of post-flare loops. Although the results we present require more quantitative validation, they constitute a prime example of how nonthermally accelerated test {electrons} can be employed to construct the observable signals in post-flare loop modeling. Our approach also leaves ample ground for improvements, as we describe below, which we will pursue in future work.

First, in our MHD setup we are only considering an isolated looptop source without the reconnection region above. We are therefore neglecting the interaction of the reconnection exhaust with the looptop structure, which could produce drastically different dynamics involving large-scale shocks and subsequent particle acceleration. We also remark the fact that HXR production in our looptop source should be eventually compared with the HXR emission from the reconnection region, which may completely outshine the looptop. This may be case-dependent, and a larger and more complex set of simulations, where the reconnection region is included, is needed to address the issue. In that context, we could start from the recent 3D MHD simulations of \cite{ruan2023}, where it was established that volume-filling turbulence, induced by KH processes related in the reconnection outflows, will develop during the impulsive phase.

A second point requiring attention stems from our particle modeling. Assuming the test-particle approach is correct (i.e.\ that high-energy, nonthermal populations contain a negligible fraction of the looptop plasma energy), it remains to be evaluated how important kinetic effects discarded by the GCA paradigm could be. For example, high-energy {electrons} could attain Larmor radii large enough to interact with electromagnetic fluctations over relatively large scales, which would produce an additional energization channel we are not considering. Simulating particle ensembles with the full equations of motion is extremely expensive, and needs to be left for future work if it is at all possible. In addition, it is known that particle simulations in 2D geometries (with an ignorable third direction) are affected by numerical artifacts potentially quenching cross-field-line particle motion (\citealt{jokipii1993,jones1998}). {The problem of dimensionality also affects the development of turbulence itself, which broadly presents qualitative differences in 2D and 3D. However, it has been proven that, in terms of particle acceleration, 2D and 3D turbulence produce very similar results (e.g.\  \cite{comissosironi2018,comissosironi2019}), albeit in different (close to relativistic) energy regimes. We therefore expect no dramatic differences in our results when we explore 3D setups, but this will need to be assessed. In summary, further work is needed to verify our results with 3D simulations, to ensure that the particle-trapping effects we quantified in this paper carry over to more realistic 3D simulations.}

Finally, our MHD model requires thorough evaluation before quantitative statements can be made. {First, our post-flare model currently does not include particle injection via reconnection above the looptop, which will be included in future work with more refined flare models (e.g.\ \citealt{druett2023AA,druett2023SP}).} Moreover, the quality of the MHD turbulence in the looptop, in terms of how well the energy cascade can be captured, is entirely dependent on dissipation terms and numerical resolution. Here, we have excluded explicit dissipation (e.g.\ resistivity), implying that there is no intrinsic dissipative scale in our turbulence, and dissipation is instead driven on the numerical grid by truncation errors. When changing the resolution then, the MHD results are expected to change, and the same should apply to the particle dynamics. In Appendix~\ref{app:resolution} we particularly elaborate on the effect of resolution on test-particle energization. However, the lack of explicit resistivity is perhaps the most crucial point, because without parallel electric fields our test {electrons} are essentially insensitive to impulsive acceleration (e.g.\ at reconnection sites) in the parallel direction. It has been recently argued in other contexts that particle acceleration in plasma turbulence is a multi-stage process, where particle ``injection'' occurs by means of parallel electric fields (\citealt{comissosironi2019,guo2019,zhdankin2019}). The injected {electrons} then continue accelerating via ideal-MHD processes. Here, we can only possibly model the latter, while resistivity (which would allow for parallel electric fields) is absent. In the future, we will quantify the importance of injection by conducting nonideal-MHD runs including resistivity.

In summary, our work opens several possible pathways to further investigate the dynamics of {electrons} in macroscopic coronal structures such as turbulent loops. We have demonstrated the possibility of constructing observables from MHD augmented with particle information, which takes us a step further toward the self-consistent modeling of the solar corona with first-principles methods.

\section*{Acknowledgements}
F.B.\ would like to thank Tom Van Doorsselaere and Malcolm Druett for useful discussions throughout the development of this work.
The computational resources and services used in this work were provided by the VSC (Flemish Supercomputer Center), funded by the Research Foundation -- Flanders (FWO) and the Flemish Government -- department EWI.
F.B.\ acknowledges support from the FED-tWIN programme (profile Prf-2020-004, project ``ENERGY'') issued by BELSPO, and from the FWO Junior Research Project G020224N granted by the Research Foundation -- Flanders (FWO).
R.K.\ is supported by Internal Funds KU Leuven through the project C14/19/089 TRACESpace and an FWO project G0B4521N.
R.K.\ and W.R.\ acknowledge funding from the European Research Council (ERC) under the European Union Horizon 2020 research and innovation program (grant agreement No.\ 833251 PROMINENT ERC-ADG 2018).

\section*{Data Availability}
The data underlying this article will be shared on reasonable request to the corresponding author.







\appendix

\section{Effect of numerical resolution on test-particle energization}
\label{app:resolution}

Because of our ideal-MHD model, the macroscopic properties of turbulence inside the looptop are heavily dependent on the numerical resolution. More specifically, the turbulent cascade can progress down to the smallest scales resolved in the simulation, i.e.\ the grid spacing of the highest AMR level. There, turbulent energy is dissipated numerically; by increasing the numerical resolution, one can in principle allow the cascade to progress to smaller scales, and the corresponding inertial range to extend indefinitely. For this reason, it is not expected to find convergence in the properties of turbulence by simply increasing numerical resolution above a certain threshold. On the contrary, in viscous-/resistive-MHD simulations the dissipation scales are set by an explicit viscosity/resistivity, and it is possible to obtain converged results by resolving the dissipation scales on the grid (e.g.\ \citealt{ripperda2020}).

\begin{figure}
\centering
\includegraphics[width=\columnwidth, trim={0mm 0mm 0mm 0mm}, clip]{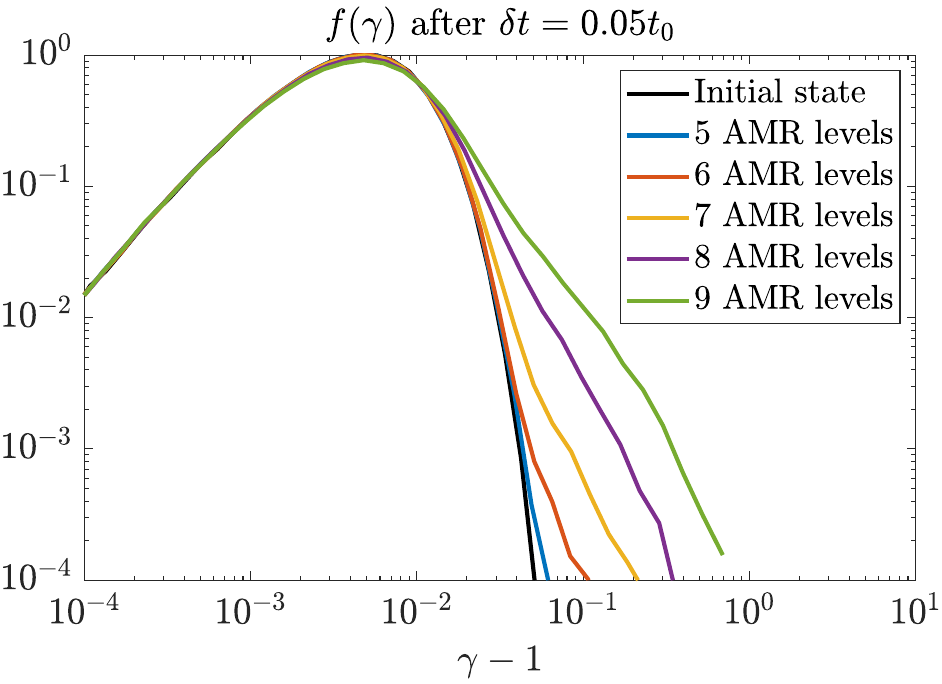}
\caption{Evolution of particle energy distributions during $t\in[2.5,2.55]t_0$ when employing different resolutions in the MHD background, from 5 to 9 AMR levels. The initial Maxwellian evolves to different states, depending on the resolution, within this short time frame.}
\label{fig:res_study}
\end{figure}

Since particle energization in our simulations depends on the properties of turbulence, by the argument above our test-particle results are expected to depend on the numerical resolution. In particular, we expect scattering mechanisms to be more efficient when the turbulent cascade progresses closer to the kinetic scales, because magnetic-field fluctuations can exist over a wider range of spatial scales. To test the effect of numerical resolution, we have performed short simulations (for a duration $\delta t=0.05t_0$) with the same initial conditions mentioned in Sec.~\ref{sec:trapping} and \ref{sec:energization}, increasing the number of AMR levels from 5 up to 9. The results are shown in Fig.~\ref{fig:res_study}, where we plot the energy distribution of all particles inside the looptop at the end of the run, compared to the initial Maxwellian distribution. Even within such a short time period, evident differences in the energy distribution develop between runs with different numerical resolutions. The distribution becomes progressively more energetic as the resolution is increased, and the results do not appear to converge even with many AMR levels (as expected from the argument above).

An interesting question is whether the difference in energy distribution is actually secular, or if it is only the result of different energization rates. If, for example, the time evolution of the energy distribution reaches a common, resolution-agnostic steady state after some time, it is possible that higher numerical resolutions simply push particles to this state faster. If, on the other hand, the final state also qualitatively depends on the numerical resolution, matters become more complicated. In such a case, it could become difficult to make definitive statements on the physics of energization processes in ideal-MHD simulations. However, testing these possibilities would require running large simulations (even larger than those presented here) for long times (i.e.\ until the distributions have converged), which is beyond our current possibilities.


\bsp	
\label{lastpage}
\end{document}